\documentclass[preprint,prd,showpacs,preprintnumbers,superscriptaddress,floatfix]{revtex4}

\usepackage{amsmath}
\usepackage{amssymb}
\usepackage{graphicx}
\usepackage{enumerate}
\usepackage{bibentry}
\usepackage{dcolumn}

\def\pslash{\rlap{\hspace{0.02cm}/}{p}}
\def\qslash{\rlap{\hspace{0.02cm}/}{q}}

\def\Dslash{\rlap{\hspace{0.07cm}/}{D}}

\def\Tr{\rm Tr\,}

\def\LLp{\textit{LLp}}
\def\LLc{\textit{LLc}}

\begin{document}
%----------------------------------------------------------------

%================================================================

\title{
Eighth-Order Vacuum-Polarization Function Formed by \\
Two Light-by-Light-Scattering Diagrams\\
and its Contribution
to the Tenth-Order Electron $g\!-\!2$
}

%================================================================

\author{T.~Aoyama} 
\affiliation{Institute of Particle and Nuclear Studies, 
High Energy Accelerator Research Organization (KEK), Tsukuba, 
305-0801, Japan} 

\author{M.~Hayakawa} 
\affiliation{Department of Physics, Nagoya University, Nagoya, 
Aichi 464-8602, Japan} 

\author{T.~Kinoshita} 
\affiliation{Laboratory for Elementary-Particle Physics, 
Cornell University, Ithaca, New York 14853, U.S.A.}  

\author{M.~Nio} 
\affiliation{Theoretical Physics Laboratory, Nishina Center, RIKEN, Wako, 
351-0198, Japan} 

\author{N.~Watanabe} 
\email{noriaki@eken.phys.nagoya-u.ac.jp}
\affiliation{Department of Physics, Nagoya University, Nagoya, 
Aichi 464-8602, Japan}

%================================================================

%================================================================

\begin{abstract} 
We have evaluated the contribution to the  anomalous magnetic
moment of the electron from six tenth-order Feynman diagrams 
which contain eighth-order vacuum-polarization function 
formed by two light-by-light scattering diagrams connected
by three photons. The integrals are constructed 
by two different methods.  
In the first method the subtractive counter terms are used to deal with 
ultraviolet (UV) singularities together with the requirement
of gauge-invariance.
In the second method, the Ward-Takahashi identity is applied to 
the light-by-light scattering amplitudes to eliminate UV singularities. 
Numerical evaluation confirms that the two methods
are consistent with each other
within their numerical uncertainties. 
Combining the two results statistically and adding small contribution from
the muons and/or tau leptons, we obtain $ 0.000~399~9~(18)~(\alpha/\pi)^5$.
We also evaluated the contribution to the muon $g\!-\!2$ 
from the same set of diagrams and found $ -1.263~44~(14)~(\alpha/\pi)^5$.  

\end{abstract}


\maketitle 

%----------------------------------------------------------------

\section{Introduction}
\label{sec:intro}

The anomalous magnetic moment ($g\!-\!2$) of the electron has played the
central role in testing the validity of quantum electrodynamics (QED)
since its experimental and theoretical discovery in 
1940's \cite{Kusch:1947,Schwinger:1948iu}.

The precision of $g\!-\!2$ measurements has been 
improved steadily in subsequent sixty years
\cite{Rich:1972, VanDyck:1987ay}. The Harvard group recently succeeded in
measuring the $g$ value of the electron with
a substantially reduced uncertainty by using a cylindrical
Penning trap. Their measurements published in 2006 \cite{Odom:2006gg} and
in 2008 \cite{Hanneke:2008tm} are
\begin{align}
  &a_e({\rm HV06}) = 1\ 159\ 652\ 180.85\ (76) \times 10^{-12} 
  \quad \quad [0.66\,{\rm ppb}] \, ,  
\label{eq:aHV06}
\\ 
 &a_e({\rm HV08}) = 1\ 159\ 652\ 180.73\ (28) \times 10^{-12} 
  \quad \quad [0.24\,{\rm ppb}] \, . 
\label{eq:aHV08}
\end{align}

Taking the presence of the muon and tau lepton into account
the QED contribution to the electron $g\!-\!2$ can be written
in the general form
\begin{equation}
a_e (\text{QED}) = A_1 + A_2 (m_e/m_\mu) + A_2 (m_e/m_\tau)
 + A_3 (m_e/m_\mu,m_e/m_\tau),
\end{equation}
where $A_i$ can be expanded into power series in $\frac{\alpha}{\pi}$
\begin{equation}
A_i = A_i^{(2)} \left ( \frac{\alpha}{\pi} \right )
     + A_i^{(4)} \left ( \frac{\alpha}{\pi} \right )^2
     + A_i^{(6)} \left ( \frac{\alpha}{\pi} \right )^3
     + \ldots, \qquad i=1, 2, 3,
\end{equation}
whose coefficients are finite calculable quantities, 
which is guaranteed by the renormalizability of QED.
Thus far the coefficients up to the eighth-order have been calculated 
\cite{Petermann:1957,Sommerfield:1957,Kinoshita:1995,Laporta:1996mq,
Kinoshita:2005zr,Aoyama:2007dv,Aoyama:2007mn,
Samuel:1990qf,Li:1992xf,Czarnecki:1998rc,%
Laporta:1993ju,Laporta:1992pa,%
Lautrup:1977hs,%
Passera:2006gc}.
The small but non-negligible corrections due to  hadrons  
\cite{Davier:1998si,Krause:1996,Melnikov:2003xd,Bijnens:2007pz} 
and weak interactions \cite{Czarnecki:1996ww} are also known 
with sufficient precision.

Combining the
experiment and the theory,  one can determine the value
of the fine structure constant $\alpha$ 
\cite{Gabrielse:2006gg,Gabrielse:2006ggE,Hanneke:2008tm}
\begin{equation}
 \alpha^{-1}(a_e) = 
  137.035~999~084~(12)(37)(33)  
  \quad \quad [0.37\,{\rm ppb}] \, , 
 \label{eq:alphaFromg-2}
\end{equation} 
where the uncertainties come from  numerical errors in the eighth-order term 
\cite{Aoyama:2007dv,Aoyama:2007mn},
an educated guess of the tenth-order term 
\cite{Mohr:2005zz}, and the experiment (\ref{eq:aHV08}), in that order.
Note that, for the first time in three decades,
the experimental uncertainty ($0.33 \times 10^{-7}$) 
has been reduced to a value smaller than
the combined theoretical uncertainty ($0.39 \times 10^{-7}$).
The uncertainty of this $\alpha$ is  about 20 times smaller than those of other
independent methods, such as a Rb recoil velocity determination in an optical
lattice \cite{Clade:2006PRA} or a Cs recoil velocity in an atom interferometry 
\cite{Wicht:2002, Gerginov:2006}.  
A new  Cs measurement 
is now in progress, which is designed to obtain the value of $\alpha$ 
with  the relative uncertainty   $0.3$ ppb \cite{Mueller:2008b}.
Such forthcoming progress of the atomic physics experiments 
will enable us to check 
the validity of QED %quantum theory of electrodynamics 
with the accuracy less than $0.1$ ppb  
by examining 
consistency of various values of $\alpha$. 

Turning back to the electron $g\!-\!2$, we find that
the largest theoretical uncertainty now comes from the tenth-order term 
$A_1^{(10)}$.
Clearly an actual value, not an estimate, of this term
is urgently needed.   
There are 12672 Feynman diagrams contributing to $A_1^{(10)}$.
Our on-going effort to evaluate all of them has been reported 
in several articles 
\cite{Kinoshita:2005sm,Aoyama:2005kf,Aoyama:2007bs,Kinoshita:2006kh,
Aoyama:2006hg,Kinoshita:2006az,Nio:2007zz,Aoyama:2007RD}.
In this paper, we report the contribution from the diagrams belonging to the 
gauge-invariant set Set I(j).
These diagrams contain 
the eighth-order vacuum-polarization diagram formed by two
light-by-light scattering diagrams connected by three photons,
which was constructed first time in this work. 
Although the Set I(j) consists of only six Feynman diagrams
and it turns out to be numerically very small,
it has features not found in other 12666 diagrams contributing 
to the tenth-order electron $g\!-\!2$.
Thus it deserves a special treatment as is described in this paper.

The primary purpose of this paper is to report
the contribution of the gauge-invariant set Set I(j) to the 
mass-independent term $A_1^{(10)}$ of the electron $g\!-\!2$.
The contributions to $A_2^{(10)}$ from closed loops of electrons, muons
and/or tau leptons are  evaluated and  
reported separately in Sec.~\ref{sec:numerical}.
Summing up all contributions, we obtained the tenth-order contribution from
Set I(j)  
\begin{align}
a_e^{(10)}(\text{Set I(j)}) 
          &=  \left ( A_1^{(10)} (\text{Set I(j)})
                     +A_2^{(10)} (m_e/ m_\mu ) ( \text{Set I(j)})   
                                \right ) \left ( \frac{\alpha}{\pi} \right ) ^5  
                                \nonumber \\   
          &=  0.000~399~9 ~ (18)~ \left ( \frac{\alpha}{\pi} \right )^5~.
\label{eq:electron-final-result}
\end{align} 
The contribution from the tau lepton is smaller than the uncertainty quoted here.

The contribution of Set I(j) 
to the muon $g\!-\!2$ can be obtained by replacing
the external (or open) 
electron line by a muon line, keeping the internal fermion loops intact.
The result of numerical integration gives the mass-dependent term of the
muon $g\!-\!2$ 
\begin{align}
a_\mu^{(10)}(\text{Set I(j)})
       &= 
         \left ( A_2^{(10)} (m_\mu/ m_e ) (\text{Set I(j)}) 
               + A_2^{(10)}(m_\mu/m_\tau) (\text{Set I(j)})
          \right .       
                              \nonumber \\
       & \qquad \left .
        + A_3^{(10)} (m_\mu/m_e, m_\mu /m_\tau ) ( \text{Set I(j)} )
         \right ) \left ( \frac{\alpha}{\pi} \right ) ^5  
                                \nonumber \\
       &=   -1.263~44~(14)~ \left ( \frac{\alpha}{\pi} \right ) ^5~.
\label{eq:muon-final-result}     
\end{align}
The contribution from tau lepton is of order of the uncertainty quoted here.

In Sec.~\ref{sec:method} we describe how to construct the eighth-order vacuum-polarization
function of Set I(j). Three possible ways are considered. In Sec.~\ref{sec:detail}
two of three methods
are described in detail. The utility of the Ward-Takahashi identity
applied to a vacuum-polarization diagram and a light-by-light scattering diagram
is particularly emphasized. 
Once the vacuum-polarization function is constructed, its contribution to the
tenth-order anomaly is easily calculated.
The details of the numerical results are presented in Sec.~\ref{sec:numerical}.
Sec.~\ref{sec:conclusion} is devoted to conclusion and discussion.
Appendix~\ref{AppendixA} 
describes new features of the vacuum-polarization function
for the diagrams of Set I(j) and also shows how to obtain
its contribution to the magnetic moment which does not
rely on the photon spectral function explicitly.
An example of the structure of the
integrand used in the Method C is shown in
Appendix~\ref{AppendixB}.

\begin{figure}[t]
\includegraphics[width=15cm]{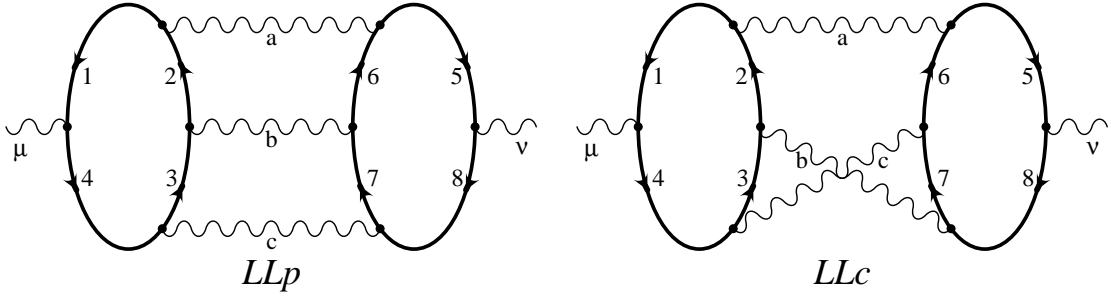}
\caption{Eighth-order vacuum-polarization diagrams \LLp\ and \LLc. 
There are two diagrams of \LLp\ type and four diagrams of \LLc\ type.
When inserted into a photon line of the second-order vertex diagram, 
they give the tenth-order
diagrams contributing  to the lepton $g\!-\!2$  called Set I(j).
}
\label{fig:LLp&LLc}
\end{figure}
%----------------------------------------------------------------

\section{Eighth-order vacuum-polarization diagrams which consist of two light-by-light-scattering subdiagrams 
connected by three photons}
\label{sec:method}

Two approaches are found in the literature for dealing with
the insertion of a vacuum-polarization diagram in the 
photon line of the second-order vertex diagram.
One is to take advantage of the fact that such an insertion
amounts to replacing the photon line by a sum of massive vector particles
weighted by the spectral function, which is the absorptive part
of the vacuum-polarization function.
Another is to insert the vacuum-polarization function itself
obtained by the Feynman-Dyson rules.
The first method is very convenient if the spectral function is known
exactly \cite{Hoang:1995ex}, or in good approximation \cite{Broadhurst:1992za}.
In most cases where such a spectral function is not available, however,
one is forced to choose the second approach.
The tenth-order diagrams of Set I(j),
which consist of eighth-order vacuum-polarization functions
inserted into the second-order vertex diagram,
belong to the latter.
This approach was initially developed in 
Refs.~\cite{Kinoshita:1979dy,Kinoshita:1979ej}.
(See also Eqs.~(5.6) and~(5.8) of Ref.~\cite{Kinoshita:1990}).
A more transparent and compact form is presented in
\cite{Broadhurst:1992si}:
\begin{align} 
a^{(2+n)} = 
  - \int_0^1 dy (1-y)\, 
     \Pi^{(n)}\left (-\frac{y^2 }{1 - y} \right ) \, , 
   \label{eq:fromBroadhurst} 
\end{align} 
where $a^{(2+n)}$ stands for the $(2+n)$th-order electron anomaly that
is obtained from the second-order vertex diagram in which  
the renormalized $n$th-order vacuum-polarization function $\Pi^{(n)}$ 
is inserted.  
A derivation of Eq.~(\ref{eq:fromBroadhurst}) is given in Appendix~\ref{AppendixA}.

In the second approach the problem is thus reduced to
an explicit construction of $\Pi$ 
from  the gauge-invariant set Set I(j) of Feynman diagrams.
When twisted and flipped appropriately, two
of the vacuum-polarization diagrams of Set I(j) (called \LLp) 
are reduced to planar form with three uncrossed photons, and
four of them  (called \LLc) have lower two of the photon lines crossed
(see Fig.~\ref{fig:LLp&LLc}).
Applying Feynman-Dyson rules formally to one of the \LLp-type diagrams 
we obtain
\begin{multline}
  \Pi_{\LLp}^{\mu\nu} (q)
  = (-1)^2 \frac{1}{(2\pi)^8} \left(\frac{\alpha}{\pi}\right)^4 
  \int\!d^4 l_1 \int\!d^4 l_2 \int\!d^4 l_3 \int\!d^4 l_4 \\
  \times \Tr\left[
    \gamma^\mu  
    \frac{1}{\pslash_1-m}
    \gamma^\alpha 
    \frac{1}{\pslash_2-m} 
    \gamma^\beta 
    \frac{1}{\pslash_3-m}
    \gamma^\zeta 
    \frac{1}{\pslash_4-m}
    \right]
  \frac{1}{p_a^2}\,\frac{1}{p_{b}^2}\,\frac{1}{p_{c}^2} \\
  \times \Tr\left[
    \gamma^\nu  
    \frac{1}{\pslash_5-m}
    \gamma_\alpha 
    \frac{1}{\pslash_6-m} 
    \gamma_\beta 
    \frac{1}{\pslash_7-m}
    \gamma_\zeta 
    \frac{1}{\pslash_8-m}
    \right],
\label{llp} 
\end{multline}
where each closed lepton loop contributes a factor $-1$, 
$p_i$ are linear combinations of loop momenta $l_1$, $l_2$, $l_3$, $l_4$ 
and external momentum $q$, which enters
at the $\mu$ vertex and leaving at the $\nu$ vertex (see Fig.~\ref{fig:LLp&LLc}).

The second \LLp-type diagram is obtained by reversing the direction
of the arrow of lepton lines in the second trace of Eq.~(\ref{llp}).
By charge-conjugation invariance of QED
it is equivalent to the first one.
The \LLc-type diagrams are obtained by exchanging 
$\gamma_\beta$ and $\gamma_\zeta$ in the second trace of Eq.~(\ref{llp}).
All four diagrams of \LLc-type are equivalent to each other.

Of course formal 
expressions such as Eq.~(\ref{llp}) are UV-divergent and meaningless
until they are regularized properly.
We follow the standard procedure to extract physical information
from the expression (\ref{llp}) and a similar one for \LLc:

\begin{enumerate}[(i)]
\item  Make them convergent by the Pauli-Villars regularization
of lepton loops and the Feynman cutoff of photon propagators.

\item Renormalize them by subtractive renormalization,
where subtraction integrals 
must be regularized in the same way as in (i).

\item Remove the regularization terms from the final renormalized formula.
\end{enumerate}

\vspace{5mm}
\noindent
These steps ensure that individual integrals obtained are finite. 
However they still contain terms which are not gauge-invariant.
These terms cancel out only after they are 
summed over the gauge-invariant set of Feynman diagrams.
Some details of the steps are described in the following.

\begin{figure}[t]
\includegraphics[width=15cm]{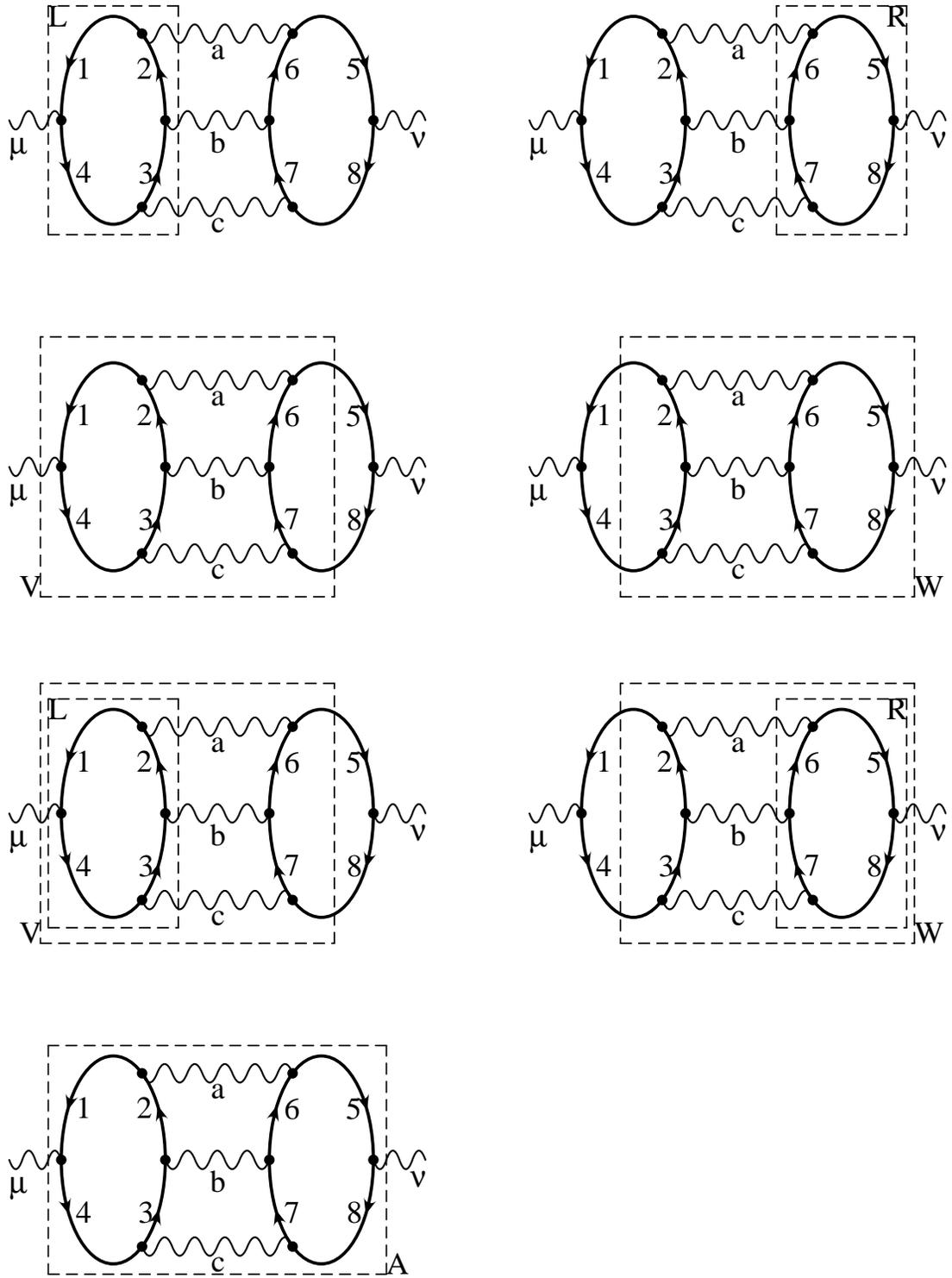}
\caption{Divergence structure of \LLp. 
Subdiagrams are $L$, $R$, $V$, $W$, $L$ in $V$, and $R$ in $W$, and the
whole diagram $A$ of \LLp.}
\label{fig:LLp_uv}
\end{figure}

The integral (\ref{llp})
has eight UV-divergent subdiagrams, including itself.
They are, namely, 
the light-by-light-scattering subdiagram $L$ (formed by a closed loop
of lepton lines $1$, $2$, $3$, $4$), another
light-by-light-scattering subdiagram $R$ (formed by a closed loop
of lepton lines $5$, $6$, $7$, $8$), 
a sixth-order vertex diagram $V$ (formed by lepton lines 
$1$, $2$, $3$, $4$, $6$, $7$ and photon lines $a$, $b$, $c$),
another sixth-order vertex diagram $W$ (formed by lepton lines
$2$, $3$, $5$, $6$, $7$, $8$ and photon lines $a$, $b$, $c$),
diagrams of the type ($L$ in $V$) and ($R$ in $W$),
and the diagram itself that consists of all
lepton lines $1$, $2$, $3$, $4$, $5$, $6$, $7$, $8$ and 
all photon lines $a$, $b$, $c$ (see Fig.~\ref{fig:LLp_uv}). 
One more type of UV-divergence caused by $L$ and $R$ together
generates no terms which contribute to the anomaly.
Thus it can be ignored.

UV divergences coming from $L$ and $R$ 
are only logarithmic and
can be controlled by the Pauli-Villars regularization
of the lepton loop.
Control of UV divergences of $V$ and $W$ 
requires Pauli-Villars regularization
as well as Feynman cut-off of virtual photon momenta. In the latter, 
the photon propagator with momentum $k$ is regularized as
\begin{equation}
\frac{ 1 }{ k^2 - \lambda^2 } \rightarrow 
            - \int_{\lambda^2}^{\Lambda^2} dM^2 \frac {1 }{ (k^2 - M^2)^2 }~,
\label{Feynman-regularization}
\end{equation}   
where the photon mass $\lambda$ and the UV cut-off $\Lambda$ are introduced 
temporarily 
and  to be put to zero and infinity, respectively,  
in the end.
Finally, we must control the UV divergence involving 
all lepton lines and all photon lines.
It is important to note that this divergence cannot be controlled by
Pauli-Villars regularizations of two closed lepton loops alone.
The quadratic behavior of this divergence comes mostly from
three photons working together, a novel feature encountered
for the first time in the eighth-order vacuum polarization.

The sum $\Pi^{\mu\nu} (q)$ of all six diagrams, two of
\LLp\ type and four of \LLc\ type, is gauge invariant
and completely free of divergence after charge renormalization is
carried out.  However,
in our numerical work which adopts
the parametric integral formulation
based on the topology of an individual Feynman diagram
\cite{Cvitanovic:1974uf,Cvitanovic:1974sv}, 
it is more convenient to deal with the diagrams \LLp\ and \LLc\ separately.
This means that we must go one step backwards and
explicitly carry out the 
renormalization of logarithmic divergence from light-by-light 
scattering subdiagrams, etc., as well as the quadratic divergence from 
the vacuum-polarization diagram as a whole. 
The logarithmic divergence is very mild and 
its removal by renormalization can be handled 
within the numerical framework keeping the gauge-invariance rigorously.

The standard way to handle the quadratic UV divergence 
is to note that the Lorentz covariance dictates that 
$\Pi_\mathcal{G}^{\mu\nu}$, of either $\mathcal{G}=$\LLp\ or \LLc\ ,
consists of two scalar functions $\Pi_\mathcal{G}^{(a)}(q^2)$ and $\Pi_\mathcal{G}^{(b)}(q^2)$:
\begin{equation}
  \Pi_\mathcal{G}^{\mu\nu} (q)
  =  g^{\mu\nu}\,\Pi_\mathcal{G}^{(a)}(q^2) 
    + q^\mu q^\nu\,\Pi_\mathcal{G}^{(b)}(q^2) \, , 
\end{equation}
and note that $\Pi_\mathcal{G}^{\mu \nu}(q)$ has 
the dimension of square of momentum so that
the quadratic  divergence (which is proportional to the cut-off
momentum squared) is confined to
the term proportional to $g^{\mu \nu}$, 
or more precisely to the $q$-independent part of $\Pi_\mathcal{G}^{(a)}(q^2)$.
Thus, if we write $\Pi_\mathcal{G}^{(a)} (q^2)$ as
\begin{equation}
\Pi_\mathcal{G}^{(a)}(q^2) = [\Pi_\mathcal{G}^{(a)}(q^2) - \Pi_\mathcal{G}^{(a)}(0) ]
 + \Pi_\mathcal{G}^{(a)}(0),
\label{regularizedPi}
\end{equation}
the term within the parentheses is free from the quadratic divergence.

As is well-known, gauge-invariance dictates that
quadratic divergences in $\Pi_\mathcal{G}^{\mu\nu}(q)$ of the
individual diagrams 
should disappear from the sum of the gauge-invariant set of the diagrams,  and 
$\Pi^{\mu\nu}(q)\equiv 4 \Pi_\LLp^{\mu \nu}+ 2 \Pi_\LLc^{\mu \nu}$ should satisfy 
the transversality condition
\begin{equation}
\Pi^{\mu \nu}(q) = ( q^\mu q^\nu - q^2 g^{\mu \nu} ) \Pi^{(b)} (q^2).
\label{vacpolf}
\end{equation}
The scalar function $\Pi^{(b)} (q^2)\equiv2\Pi_\LLp^{(b)}(q^2) + 4 \Pi_\LLc^{(b)}(q^2)$ 
defined by this equation
is free from all subdiagram UV divergences.
However it still has an overall UV divergence which
must be removed by subtraction of
$\Pi^{(b)} (0)$, which is nothing but charge renormalization.

These observations lead us to three possible methods for 
obtaining the renormalized (not yet gauge invariant)
amplitude $\Pi_\mathcal{G} (q^2)$, which is the \LLp\ or \LLc\ part
of $\Pi^{(b)} (q^2) - \Pi^{(b)}(0)$.
The first method is

\vspace*{5mm}
\noindent
\begin{minipage}{6in}
{\bf Method A. Collect all terms of $\Pi^{\mu\nu}(q)$ 
which are coefficients of $q^\mu q^\nu$.}
\end{minipage}

\vspace*{5mm}
Another approach is to note that Eq.~(\ref{vacpolf}) implies 
\begin{equation}
q_\lambda \Pi^{\lambda\nu}(q) = 0,
\label{gauge}
\end{equation}
which is valid for arbitrary $q$.
Differentiating this equation with respect to $q_\mu$ we
obtain
\begin{equation}
 \Pi^{\mu\nu} (q) = - q_\lambda \frac{\partial}{\partial q_\mu} \Pi^{\lambda\nu}(q),
\label{vacpolf2}
\end{equation}
in which one power of $q$ is extracted explicitly.
This has the effect of removing the quadratic UV divergence automatically.
Thus, we can choose 

\vspace*{5mm}
\noindent
\begin{minipage}{6in}
{\bf Method B. Collect coefficients of $q^\mu q^\nu$ or those of
$-g^{\mu\nu} q^2$ of 
$\Pi^{\mu\nu}$ from the right-hand side of Eq.~(\ref{vacpolf2}).}
\end{minipage}

\vspace*{5mm}
Yet another approach is to start from
the equation involving the second derivative of $\Pi^{\mu\nu} (q)$:
\begin{equation}
 \Pi^{\mu\nu} (q) = \frac{1}{2}  q_\lambda q_\sigma \frac{\partial}{\partial q_\mu} 
\frac{\partial}{\partial q_\nu} \Pi^{\lambda\sigma}(q),
\label{vacpolf3}
\end{equation}
which follows from Eq.~(\ref{vacpolf2}) and 
\begin{equation}
q_\lambda q_\sigma \Pi^{\lambda\sigma}(q) = 0,
\label{gauge2}
\end{equation}
and symmetry of $\Pi^{\mu\nu}$ in $\mu$ and $\nu$.
Thus we may also start from the following rule
in which two powers of $q$ are extracted explicitly:

\vspace*{5mm}
\noindent
\begin{minipage}{6in}
{\bf Method C. Collect coefficients of $q^\mu q^\nu$ or those of
$-g^{\mu\nu} q^2$ of 
$\Pi^{\mu\nu}$ from the right-hand side of Eq.~(\ref{vacpolf3}).}
\end{minipage}
\vspace*{5mm}

It turns out that Method C has a distinct advantage over the other two. 
Not only the quadratic divergence but also 
subdiagram logarithmic UV divergences,
except for the one requiring charge renormalization,
are eliminated as a consequence of the second derivative. 
Aside from this difference, however, Method A and Method B
are equally useful and effective 
as Method C for carrying out numerical evaluation of 
the contribution of the Set I(j).

%-----------------------------------------------------------------

\section{Construction of the vacuum-polarization function $\Pi (q^2)$}
\label{sec:detail}

\subsection{Parametric representation of $\Pi (q^2)$ }
\label{sec:regularization}

Once the integral is made finite by regularization, 
we can safely deal with individual diagrams.
In this article we adopt the method of parametric representation
which has been successfully applied to similar problems
\cite{Kinoshita:1990}.
We begin by replacing
the numerator, e.g., of Eq.~(\ref{llp}), by an operator
\begin{eqnarray}
{\cal F}^{\mu\nu} &\equiv & \Tr [ \gamma^\mu  
(\Dslash_1+m)
\gamma^\alpha 
(\Dslash_2+m) 
 \gamma^\beta 
(\Dslash_3+m)
 \gamma^\zeta 
(\Dslash_4+m)
] \nonumber \\
&\times& \Tr [ \gamma^\nu 
(\Dslash_5+m)
\gamma_\alpha 
(\Dslash_6+m) 
 \gamma_\beta 
(\Dslash_7+m)
 \gamma_\zeta 
(\Dslash_8+m)
],
\label{numop} 
\end{eqnarray}
where
\cite{Karplus:1950}
\begin{equation}
D_j^\mu = \frac{1}{2} \int_{m_j^2}^\infty dm_j^2 \frac{\partial}{\partial q_{j\mu}},
\label{Dop}
\end{equation}
and bring it in front of the momentum integration.
(This may not be as straightforward as it sounds,  
and requires a more careful argument
of Pauli-Villars regularization.  But the end result is correct.)
Then we combine all denominators
with the help of Feynman parameters $z_1$, \dots , $z_8$ for leptons and 
$z_a$, $z_b$, $z_c$ for photons:
\begin{eqnarray}
&&\frac{1}{p_a^2} \frac{1}{p_{b}^2} \frac{1}{p_{c}^2} 
\prod_{i=1}^8 \frac{1}{p_i^2-m_i^2}
\nonumber \\
&&=  10 ! \int (dz)  \frac{1}{(\sum_{i=1}^8 z_i (p_i^2 - m_i^2) 
+z_a (p_a^2 - \lambda^2) +z_b (p_b^2-\lambda^2) +z_c (p_c^2-\lambda^2))^{11}},
\label{Fparameters}
\end{eqnarray}
where the photon mass $\lambda$
is introduced temporarily, to be put to zero in the end.

As usual individual photon propagators may be regularized using
the Feynman cutoff (\ref{Feynman-regularization}).
Alternately one may regularize all three photons together as follows:
\begin{eqnarray}
&&\frac{1}{(\sum_{i=1}^8 z_i (p_i^2 - m_i^2) 
+z_a (p_a^2 - \lambda^2) +z_b (p_b^2-\lambda^2) +z_c (p_c^2-\lambda^2))^{11}}
\nonumber \\
&&= -11  \int_{\lambda^2}^{\Lambda^2} dM^2
\frac{z_{abc}}{(\sum_{i=1}^8 z_i (p_i^2 - m_i^2) 
+z_a p_a^2 +z_b p_b^2 +z_c p_c^2 -z_{abc} M^2)^{12}},
\label{regularizer}
\end{eqnarray}
where $\Lambda$ is the UV cutoff and $z_{abc}=z_a+z_b+z_c$.
Let us assume that such a regularization is always done.
Now we can carry out the momentum integration and obtain
\begin{eqnarray}
& &
\int\!d^4l_1
\int\!d^4l_2
\int\!d^4l_3
\int\!d^4l_4
\frac{1}{(\sum_{i=1}^8 z_i (p_i^2 - m_i^2) 
+z_a p_a^2 +z_b p_b^2 +z_c p_c^2 -z_{abc} \lambda^2)^{11}}
\nonumber  \\
          &=& -\frac{(\pi^2 i)^4}{((11-1) !)/((11-9) !)}
 \int (dz) \frac{1}{U^2 V^3},
\label{integformula3}
\end{eqnarray}
where
\begin{equation}
V = z_{1234} m^2 + z_{5678} m^2 + z_{abc} \lambda^2 - q^2 G,
\label{defV}
\end{equation}
$z_{1234} =z_1 +z_2+ z_3 +z_4$, etc., and 
\begin{equation}
G =-z_1 A_1 +z_a A_a + z_5 A_5,
\label{defG}
\end{equation}
assuming that the photon momentum 
$q$ enters the diagram at the vertex $\mu$, goes through
lepton line $1$, photon line $a$, lepton line $5$,
and exits from  the vertex $\nu$.
$A_i$ is the scalar current associated with the line $i$ \cite{Kinoshita:1990}.
Note that $A_1$ is defined assuming that the arrow of fermion line 1
is opposite to the direction of $q$, whereas the photon line $a$ and
fermion line $5$ are in the same direction as $q$ (see Fig.~\ref{fig:LLp&LLc}).
(Actually, any continuous path of $q$ is equivalent to that of (\ref{defG}),
as far as the integral is made finite by regularization.
Note that this may not be guaranteed for divergent integrals.)
$U$ is a Jacobian from the momentum space to the Feynman parameter space.
$(dz)$ stands for the eleven dimensional integration variables of  
Feynman parameters  
with the constraint that the sum of eleven Feynman parameters is unity.
Although this integral is still logarithmically divergent,
when it is regularized with the cut-off of Eq.~(\ref{regularizer}),
the ${\cal F}^{\mu\nu}$-operation can be carried out correctly.
(If necessary, we can introduce another cutoff parameter.)

Collecting all numerical factors and bringing the operator ${\cal F}^{\mu\nu}$ back 
into the integral
we obtain, for \LLp,
\begin{eqnarray}
\Pi_{\LLp}^{\mu\nu} (q)
&=& -(-1)^2 \frac{1}{(2\pi)^8} 
\left(\frac{\alpha}{\pi}\right )^4 
(10 !) 
 \frac{(\pi^2 i)^4}{((11-1) !)/((11-9) !)}
 \int (dz) {\cal F}^{\mu\nu}
 \frac{1}{U^2 V^3}
\nonumber \\
&=& -\frac{2 !}{2^8} 
\left(\frac{\alpha}{\pi}\right )^4 
 \int (dz) 
 {\cal F}^{\mu\nu} \frac{1}{U^2 V^3}.
\label{integformula4}
\end{eqnarray}
Before proceeding further we must  carry out renormalization explicitly,
following a well-established method.
We mention here only few aspects that are specific
to Set I(j).

The first point to note is
that the renormalization constants for the sixth-order vertices
$V$ and $W$ are actually zero for the gauge-invariant quantity 
$\Pi(q^2)$
by Furry's theorem since there is no ``self-energy'' diagrams
corresponding to $V$ or $W$.
However, they are nonvanishing for individual
integrals, and must be subtracted explicitly from the unrenormalized
integral.
The leading logarithmic part of such a subtraction term 
can be readily obtained by the \textit{K}-operation \cite{Kinoshita:1990},
in which the UV divergent part of the standard on-shell renormalization
constant is used. 
However, in the case of Set I(j), 
it turns out to be better to construct the exact and full on-shell 
renormalization
term which enables us to avoid the trouble of calculating the residual 
renormalization term separately.

Similarly, for the light-by-light-scattering amplitudes of $L$ and $R$,
we can avoid residual renormalization by defining
the renormalization terms 
as the standard on-shell amplitudes
defined with all its momenta {\it external} to it put to zero. 
The gauge invariant set of this light-by-light scattering amplitude
is summed up to zero, which can also be shown by calculation with the dimensional
regularization,  hence no residual renormalization is needed.

The integrals obtained by the Method A and the Method B 
can be shown to be analytically identical
using ``Kirchhoff's laws''  on junctions and loops \cite{Kinoshita:1990}.
In the following we shall therefore consider only Method B
and Method C.

\subsection{More on Method B}
\label{sec:methodB}

We are now ready to consider Method B in detail.
Let us write the integral for \LLp\  symbolically,
ignoring explicit multiple integration,  as
\begin{equation}
\Pi = LSR,
\end{equation}
where $L$ and $R$ are light-by-light-scattering diagrams
introduced previously and
$S$ stands for the set of three photons connecting $L$ and $R$.
Then, the differentiation in Eq.~(\ref{vacpolf2})
can be carried out as
\begin{equation}
\frac{\partial}{\partial q_\mu} \Pi =
\left (\frac{\partial}{\partial q_\mu} L \right ) SR
+L \left (\frac{\partial}{\partial q_\mu} S \right ) R
+LS\left (\frac{\partial}{\partial q_\mu} R \right ).
\end{equation}
The first and third terms involve differentiation of the lepton propagators
in the closed lepton loops
while the second one is differentiation of the photon propagator.
These differentiation can be carried out using the identities
\cite{Kinoshita:1990}
\begin{eqnarray}
\frac{\partial}{\partial q_\mu} \frac{1}{\pslash+\qslash-m} &=& -2D^\mu (\Dslash+m) \frac{1}{((p+q)^2-m^2)^2},  
\label{D-derivative}
\\
\frac{\partial}{\partial q_\mu} \frac{1}{(p+q)^2} &=& \frac{-2(p+q)^\mu}{(p+q)^4}~.
\end{eqnarray}
This operation gives rise to an additional denominator factor which can be handled,
for instance, as follows:
\begin{eqnarray}
&&\frac{1}{p_a^2} \frac{1}{p_{b}^2} \frac{1}{p_{c}^2} 
\frac{1}{p_1^2-m_1^2}
\prod_{i=1}^8 \frac{1}{p_i^2-m_i^2}
\nonumber \\
&& =\frac{\partial}{\partial m_1^2} 
\frac{1}{p_a^2} \frac{1}{p_{b}^2} \frac{1}{p_{c}^2} 
\prod_{i=1}^8 \frac{1}{p_i^2-m_i^2}
\nonumber \\
&&=  11 ! \int (dz)  \frac{z_1}{(\sum_{i=1}^8 z_i (p_i^2 - m_i^2) 
+z_a (p_a^2 - \lambda^2) +z_b (p_b^2-\lambda^2) +z_c (p_c^2-\lambda^2))^{12}}.
\label{Fparameters2}
\end{eqnarray}
As a consequence, in Eq.~(\ref{integformula4}),  $1/V^3$ is replaced by $-1/V^4$,
2! is replaced by 3!, $-2 D^\mu$ and $-2 (p+q)^\mu$ are multiplied by $z_1$, etc.,
and then everything is multiplied by an overall factor $-q_\lambda$.
Recall also that the direction of $q$ is chosen to be opposite to that of $p$
of the lepton line 1.
In this manner we obtain 
\begin{eqnarray}
\Pi_{\LLp}^{\mu\nu} (q)
&=& -\frac{3 !}{2^8} 
\left(\frac{\alpha}{\pi}\right )^4 
 \int (dz)\,
(+2 z_1  D_1^\mu -2 z_a D_a^\mu - 2 z_5 D_5^\mu)
\nonumber \\
&\times& \Tr [\qslash 
(\Dslash_1+m)
\gamma^\alpha 
(\Dslash_2+m) 
 \gamma^\beta 
(\Dslash_3+m)
 \gamma^\zeta 
(\Dslash_4+m)
] 
\nonumber \\
&\times& \Tr [ \gamma^\nu 
(\Dslash_5+m)
\gamma_\alpha 
(\Dslash_6+m) 
 \gamma_\beta 
(\Dslash_7+m)
 \gamma_\zeta 
(\Dslash_8+m)
]
 \frac{1}{U^2 V^4} .
\label{integformula5}
\end{eqnarray}
Performing $D$-operation on $1/V^4$ using Eq.~(\ref{Dop}),
this integral can be expressed in terms of ``building blocks"
$z_i$, $B_{ij}$, $A_i$, where $i$, $j$ are indexes for lepton and photon lines.
Of course it must be modified by various  terms
required for renormalization.

\subsection{More on Method C}
\label{sec:methodC}

Let us now consider Method C.
In this case it is more convenient to choose the graphic representation
in which all three photon lines are parallel (or, uncrossed) in $S$.
Then $L$ can be replaced by a gauge-invariant sum of six
light-by-light-scattering diagrams, which we denote as $L^\mu$ to indicate
that it contains the vertex $\mu$.  
Similarly $R$ is replaced by $R^\nu$.
The explicit form of $L^{\mu}$ of \LLp\ is given by
\begin{equation}
L^{\mu} = 2 [ \Pi^{\mu \alpha \beta \zeta}(q, -p_a, -p_b, -p_c) +
          \Pi^{\mu \beta \zeta \alpha} (q, -p_b, -p_c, -p_a) +
          \Pi^{\mu  \zeta \alpha \beta}(q, -p_c,-p_a,-p_b) ],
\end{equation} 
where  the
light-by-light scattering tensor $\Pi^{\mu \alpha \beta \zeta}(q,-p_a,-p_b,-p_c)$ is
defined by
\begin{align}
&\Pi^{\mu \alpha \beta \zeta}(q,-p_a,-p_b, -p_c) \propto
\int\!d^4 l\,
\Tr\left[
    \gamma^\mu  
    \frac{1}{\pslash_1-m}
    \gamma^\alpha 
    \frac{1}{\pslash_2-m} 
    \gamma^\beta 
    \frac{1}{\pslash_3-m}
    \gamma^\zeta 
    \frac{1}{\pslash_4-m}
    \right ]~ ,
    \nonumber \\
& p_1=l-q,~~ p_2=l-p_b-p_c,~~ p_3=l-p_c, ~~p_4 = l,
\end{align}
with the overall momentum conservation $q=p_a+p_b+p_c$.
Actually this procedure gives six identical copies of 
the original six diagrams so that the
result must be divided by 6.

The differentiations in Eq.~(\ref{vacpolf3}),
where  $\Pi^{\lambda\sigma}$ is replaced by
$ (1/6)  L^{\lambda} S R^{\sigma}$ symbolically,
can be carried out as follows:
\begin{eqnarray}
6 \frac{\partial}{\partial q_\mu} \frac{\partial}{\partial q_\nu} \Pi^{\lambda\sigma} &=&
\frac{\partial^2 L^\lambda }{\partial q_\mu \partial q_\nu} S R^{\sigma}
+ \frac{\partial L^\lambda}{\partial q_\nu}  \frac{\partial S}{\partial q_\mu} R^\sigma
+ \frac{\partial L^\lambda}{\partial q_\nu}  S \frac{\partial R^\sigma}{\partial q_\mu} 
\nonumber \\
&+& \frac{\partial L^\lambda}{\partial q_\mu}  \frac{\partial S}{\partial q_\nu}  R^\sigma
+ L^\lambda \frac{\partial^2 S}{\partial q_\mu \partial q_\nu} R^{\sigma}
+ L^\lambda  \frac{\partial S}{\partial q_\nu} \frac{\partial R^\sigma}{\partial q_\mu} 
\nonumber \\
&+& \frac{\partial L^\lambda}{\partial q_\mu}  S \frac{\partial R^\sigma}{\partial q_\nu} 
+ L^\lambda  \frac{\partial S}{\partial q_\mu} \frac{\partial R^\sigma}{\partial q_\nu} 
+ L^\lambda  S  \frac{\partial^2 R^\sigma}{\partial q_\mu \partial  q_\nu} .
\label{twoderiv}
\end{eqnarray}
Although this looks awful, 
it can be simplified greatly using Ward-Takahashi
identities that hold for the gauge-invariant sum 
$L^\lambda$ (or $R^\sigma$)  of light-by-light-scattering diagrams \cite{Aldins:1970id}:
\begin{equation}
q_\lambda L^\lambda = 0, \qquad  q_\sigma R^\sigma = 0.
\label{wtlike}
\end{equation}
Multiplying Eq.~(\ref{twoderiv}) 
with $q_\lambda q_\sigma$ and applying Eq.~(\ref{wtlike}), we obtain
\begin{eqnarray}
 \Pi^{\mu\nu} (q) &=& \frac{1}{2}  q_\lambda q_\sigma \frac{\partial}{\partial q_\mu} 
\frac{\partial}{\partial q_\nu} \Pi^{\lambda\sigma}(q),
\nonumber \\
&=& \frac{1}{12} \left ( \left ( q_\lambda \frac{\partial L^\lambda}{\partial q_\nu}\right ) S
\left ( q_\sigma \frac{\partial R^\sigma}{\partial q_\mu}\right )
+  \left ( q_\lambda \frac{\partial L^\lambda}{\partial q_\mu}\right ) S
\left ( q_\sigma \frac{\partial R^\sigma}{\partial q_\nu} \right )\right ).
\label{vacpolf4}
\end{eqnarray}

The great advantage of this equation is that the derivatives like
$\frac{\partial L^\lambda}{\partial q_\nu} $
are \textit{UV-finite} so that 
cut-offs can be safely removed and $\Pi (q^2)$ can be evaluated without
renormalization of subdiagram divergences.
Of course the overall
logarithmic UV divergence must be disposed by charge renormalization.

\begin{figure}[t]
\includegraphics[width=16cm]{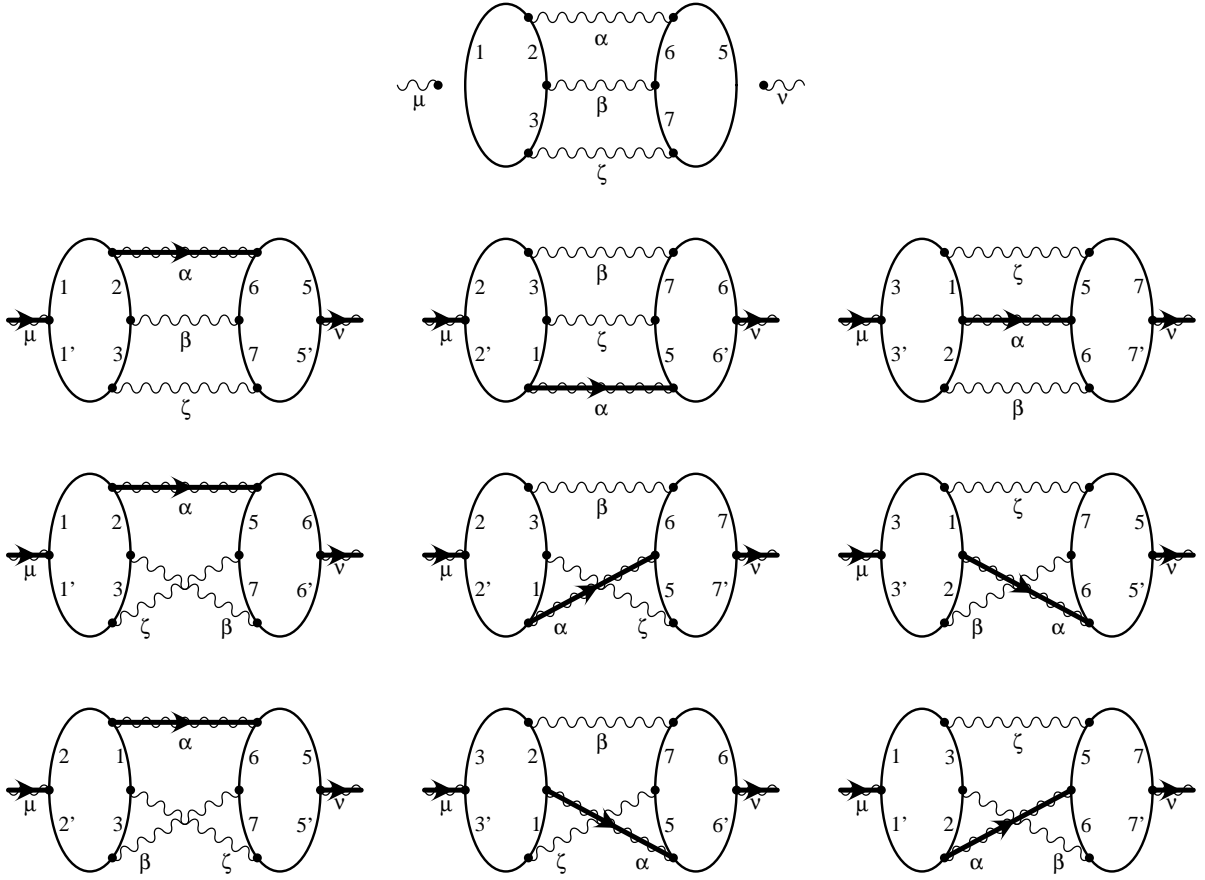}
\caption{Diagrams of $L^\mu S R^\nu$. 
Nine diagrams shown in the bottom three rows are
obtained by inserting the external photons 
labelled $\mu$ and $\nu$ into the left and right fermion loops,
respectively, of the diagram shown at the top. 
The (bold) internal photon line $\alpha$ carries the momentum $q$ in all cases.
Remaining  27 diagrams obtained by
flipping the direction of fermion loops are not shown for simplicity.}
\label{fig:LSR}
\end{figure}
    
Eq.~(\ref{vacpolf4}) provides the starting point of numerical 
evaluation by  Method C. 
To perform numerical integration, one has to decompose it 
into non-gauge-invariant forms similar to \LLp\ and \LLc. 
This can be parametrized in the same manner as for Eq.~(\ref{integformula5}),
in which light-by-light-scattering diagrams are treated as
subdiagrams of the eighth-order vacuum-polarization diagram.

In the following, however, we chose an alternate approach which emphasizes the gauge-invariant nature of
the sets $L^\mu$ and $R^\nu$ of light-by-light-scattering subdiagrams.
The set $L^\mu$ is a sum of six diagrams 
in which three photon lines and an external photon line $\mu$ 
are attached to a directed lepton loop in every possible ways. 
We prepare another set $R^\nu$ similarly.
We connect them by three photons to construct $L^\mu S R^\nu$. 
This procedure can also be stated as follows: 
suppose there is a diagram with two lepton loops connected by 
three parallel photons labelled by $\alpha$, $\beta$, $\zeta$ 
from the top to the bottom, 
and we insert an external vertex $\mu$ into the left loop, 
another external vertex $\nu$ into the right loop 
(see the top figure of Fig.~\ref{fig:LSR}). 
Disregarding the directions of lepton loops for a moment, 
there are 9 ways of insertions. 

It is found that by flipping and twisting the diagrams three of them 
are topologically equivalent to \LLp-type diagram, whose photon lines 
$a$, $b$, $c$ correspond to the cyclic permutations of 
$\alpha$, $\beta$, $\zeta$, namely, 
$\{ \alpha, \beta, \zeta \}$, 
$\{ \beta, \zeta, \alpha \}$, and 
$\{ \zeta, \alpha, \beta \}$, 
as shown in the second row of Fig.~\ref{fig:LSR}. 
Similarly, the remaining six diagrams are found to be equivalent 
to \LLc-type, whose photon lines correspond to all six permutations, 
namely, 
$\{ \alpha, \beta, \zeta \}$, 
$\{ \alpha, \zeta, \beta \}$, 
$\{ \beta, \alpha, \zeta \}$, 
$\{ \beta, \zeta, \alpha \}$, 
$\{ \zeta, \alpha, \beta \}$, and 
$\{ \zeta, \beta, \alpha \}$, 
as shown in the third and fourth rows of Fig.~\ref{fig:LSR}. 

Next we consider the flow of external momentum $q$ in the diagram. 
Three of  four external photon momenta of a light-by-light scattering diagram
are independent because of the momentum conservation.  
$L^\mu$ and $R^\nu$ in \LLp\ are connected by three photons forming 
two loops in $S$. 
Two of three
independent momenta of $L^\mu$ thus turn into two loop momenta and can be
freely shifted.  Therefore,  to fix  \textit{all} external photon momenta 
of $L^\mu$ and those of $R^\mu$ in the $L^\mu S R^\nu$ we need to 
fix  where the only one independent momentum $q$ flows in the entire
vacuum-polarization diagram.  

We define a fraction of the momentum $q_i$ flowing in the line $i$, 
and introduce a coefficient $d_i$ as 
\begin{equation}
  q_i = d_i q~.
\end{equation}
By momentum conservation, the sum of fractions flowing on three photon 
lines $a$, $b$, and $c$ of $S$ must be equal to 1: 
\begin{equation}
  d_a + d_b + d_c = 1~.
\end{equation}

Next we fix the flow of momentum through the gauge-invariant combination 
$L^\mu S R^\nu$. 
We consider the following particular choices. 

\noindent
\textit{Choice 1: $q$ flows only through the photon line $\alpha$}. 

\noindent
The combination $L^\mu S R^\nu$ is decomposed into 
\LLp-type or \LLc-type diagrams listed in Fig.~\ref{fig:LSR}, 
in which $q$ flows only on the photon labelled by $\alpha$ 
as shown by bold lines. 
For example, the diagram $\{ \alpha, \beta, \zeta \}$ corresponds 
to the \LLp-type diagram in which the fractions are given by 
$d_a = 1$, and $d_b = d_c = 0$, denoted symbolically 
as $\LLp\,(d_a = 1)$. 
Similarly, we can translate all nine diagrams of $L^\mu S R^\nu$ 
variant to \LLp\ or \LLc\ with specific values of $d_i$. 
Then, the gauge-invariant vacuum-polarization function which 
contains two sets of gauge-invariant
light-by-light-scattering subdiagrams is obtained by the combination 
\begin{align} 
  \frac{4}{6} &
  \left[
    \{ \LLp\,(d_a=1) + \LLp\,(d_b=1) + \LLp\,(d_c=1) \} 
  \right.
\nonumber \\ 
  & + 2 
  \left.
    \{ \LLc\,(d_a=1) + \LLc\,(d_b=1) + \LLc\,(d_c=1) \}
  \right],
\label{LLp-combination}
\end{align}
where it is multiplied by 4 to account for the directions of 
lepton loops and divided by 6 to take account of duplicated copies. 

Once we have selected the flow of $q$ in the photon lines, we can 
choose any flow on the fermion loops. 
For instance, we may choose the following flows for three \LLp-type diagrams: 
\[
\begin{array}{ccccc}
  \quad \text{diagram with a specific $q$-flow } \quad & ~~~~~~~~~~~~ &
  \quad \text{fermion 1--4} \quad & ~~~~~~~~~~~~ & 
  \quad \text{fermion 5--8} \quad \\
\hline
  \LLp\,(d_a=1) && d_1 = -1, && d_5 = +1, 
\\
  \LLp\,(d_b=1) && d_1 = d_2 = -1, && d_5 = d_6 = +1, 
\\
  \LLp\,(d_c=1) && d_4 = +1, && d_8 = -1,
\end{array}
\]
and other $d_i$'s are zero. 
For each diagram, different routings 
of the external momentum $q$ in the lepton lines is possible, 
but all give identical results. 
(This is nothing but a consequence of the ``Kirchhoff's laws'' 
for loops and junctions applied to \LLp \cite{Kinoshita:1990}.)

In order to drop the unwanted terms, derivatives of $S$, second derivatives of $L$ and so on,
from Eq.~(\ref{twoderiv}), we must add up the contributions from three diagrams in the
first line of Eq.~(\ref{LLp-combination}).
To do this,
we define the derivative factor $\mathbb{D}^{\mu\nu}$ 
consisting of two $D$-operators of Eq.~(\ref{D-derivative}) as 
\begin{align}
  \mathbb{D}^{\mu \nu} 
  & \equiv 
%  (-2)^2 \sum_{i=1}^{4} \sum_{j=5}^{8} d_i d_j z_i z_j D_i^{\mu} D_j^{\nu}
    \sum_{\text{three \LLp-type  diagrams}}
  \left(
    \sum_{i=1}^{4} -2 d_i z_i D_i^\mu
  \right)
  \left(
    \sum_{j=5}^{8} -2 d_j z_j D_j^\nu
  \right)
\nonumber \\
  & = 
  4 ( 
  -2 z_1 z_5 D_1^\mu D_5^\nu 
  - z_1 z_6 D_1^\mu D_6^\nu
  - z_2 z_5 D_2^\mu D_5^\nu  
  - z_2 z_6 D_2^\mu D_6^\nu 
  - z_4 z_8 D_4^\mu D_8^\nu ) \,. 
\end{align}   
This is derived by a consideration similar to the argument leading to 
Eq.~(\ref{integformula5}) from the structure of traces implicit in 
Eq.~(\ref{vacpolf4}) with the help
of Eq.~(\ref{D-derivative}). 
As seen in Fig.~\ref{fig:LSR}, $\LLp(d_c=1)$ is in fact identical with $\LLp(d_a=1)$, if
the top and bottom of the figure $\LLp(d_c=1)$ is reversed. Thus, 
we may double the contribution of $\LLp(d_a=1)$ and drop
$\LLp(d_c=1)$. Then the coefficient of $z_1z_5 D_1^\mu D_5^\nu$ becomes $-3$ and
no  $z_4z_8 D_4^\mu D_8^\nu$ term is needed.
The vacuum-polarization tensor of \LLp-type in Method C is thus given by 
\begin{multline}
  \Pi_{\LLp}^{\mu \nu} (q) 
  = -\frac{4!}{2^8} \frac{4}{6} 
  \left(\frac{\alpha}{\pi}\right )^4 
  \int (dz) \,
%  \frac{1}{2} ( \mathbb{D}^{\mu \nu} + \mathbb{D}^{\nu \mu} ) \\
  \mathbb{D}^{(\mu,\nu)} \\
  \times 
  \Tr\left[
    \qslash 
    (\Dslash_1+m)
    \gamma^\alpha 
    (\Dslash_2+m) 
    \gamma^\beta 
    (\Dslash_3+m)
    \gamma^\zeta 
    (\Dslash_4+m)
  \right] \\
  \times
  \Tr\left[
    \qslash 
    (\Dslash_5+m)
    \gamma_\alpha 
    (\Dslash_6+m) 
    \gamma_\beta 
    (\Dslash_7+m
    \gamma_\zeta 
    (\Dslash_8+m)
  \right]
  \frac{1}{U^2 V^5}\,.
\label{integformulaC}
\end{multline}
where the indexes in $\mathbb{D}^{(\mu,\nu)}$ are symmetrized with respect 
to $\mu$ and $\nu$. 
Eq.~(\ref{integformulaC}) is free from UV divergence except for the
overall charge renormalization. The Pauli-Villas regularization
is no longer required in Eq.~(\ref{integformulaC}).

Similarly, we can construct $\mathbb{D}^{\mu\nu}$ factor for \LLc: 
\begin{equation}
  \mathbb{D}^{\mu \nu} = 
  4 ( 
  - z_1 z_5 D_1^{\mu} D_5^{\nu}
  + z_1 z_8 D_1^{\mu} D_8^{\nu}
  + z_2 z_8 D_2^{\mu} D_8^{\nu}
  + z_4 z_5 D_4^{\mu} D_5^{\nu}
  + z_4 z_6 D_4^{\mu} D_6^{\nu} )~.
\end{equation}    

Another simple choice of the $q$ flow is 

\noindent
\textit{Choice 2: $\frac{1}{3}q$ flows on all internal photon lines 
$\alpha$, $\beta$, and $\zeta$.}

\noindent
In this case, all three of \LLp- (or six of \LLc-) types are 
indistinguishable. 
Thus, we find that the gauge-invariant set is 
\begin{equation}
  \frac{4}{6} 
  \left\{ 
  3 \LLp\,(d_a=d_b=d_c=1/3) 
  +  6 \LLc\,(d_a=d_b=d_c =1/3)
  \right\} \, .
\end{equation}
The $\mathbb{D}^{\mu\nu}$ operators for \LLp\ and \LLc\ with this 
choice of $q$ flow can be constructed in the same manner 
as those in \textit{Choice 1}. 
The explicit forms of the integrand thus obtained 
are different between \textit{Choice 1} and \textit{Choice 2}. 
Hereafter we shall call  Method C with \textit{Choice 1} and with \textit{Choice 2} 
as  Method C1 and Method C2, respectively.

The $D$-operators in Eq.~(\ref{integformulaC}), etc., are 
applied to the $V$-function
on the right-hand side following the ``contraction" rules \cite{Kinoshita:1990}.
Carrying out also the trace operations, the result can be written
more explicitly in the form
\begin{align} 
 \Pi_{\LLp}(q^2) 
 = 
 \left ( \frac{\alpha}{\pi} \right )^4 
& \int \frac{(dz)}{U^2} 
 \biggl  \{   
\nonumber \\  
&(H_{(1),0} + q^2 H_{(1),1} + (q^2)^2 H_{(1),2} + (q^2)^3 H_{(1),3} + (q^2)^4 H_{(1),4} ) 
                                                      \frac{1}{U V^4} 
\nonumber \\
+&(H_{(2),0} + q^2 H_{(2),1} + (q^2)^2 H_{(2),2} + (q^2)^3 H_{(2),3}  ) 
                                                        \frac{ 1}{U^2 V^3}
\nonumber \\
+&(H_{(3),0} +   q^2 H_{(3),1} + (q^2)^2 H_{(3),2}  )  \frac{1}{U^3 V^2}
\nonumber \\
+&(H_{(4),0} +   q^2 H_{(4),1}    )  \frac{ 1}{U^4 V}
\nonumber \\
+& \frac{ H_{(5),0}}{U^5} \ln  \left ( \frac{\Lambda}{V} \right )
\biggr  \},
\label{eq:Pi_parametricIntegral} 
\end{align} 
where the numerators $H_{(r),h}$ are expressed in terms of
``building blocks''  $B_{ij}$, $z_i$, and $A_i$,~$i,j=1,\dots, 8$ 
\cite{Kinoshita:1990, Aoyama:2005kf} and  $\Lambda$ is 
the UV cut-off in (\ref{regularizer}).
A detailed structure of $H_{(r),h}$ is presented in
Appendix~\ref{AppendixB}.
The charge renormalization can be trivially carried out, and the cut-off $\Lambda$ can 
be put to infinity. We thus obtain the renormalized vacuum-polarization function
\begin{equation}
\bar{\Pi}_{\LLp} (q^2) = \lim_{\Lambda \rightarrow \infty} 
                             ( \Pi_{\LLp}(q^2) - \Pi_{\LLp}(0))~. 
\end{equation}

It is straightforward to translate
Eq.~(\ref{eq:Pi_parametricIntegral}) into
numerical integration code in FORTRAN
by carrying out algebraic manipulation involved
with the help of FORM \cite{Vermaseren:2000nd}.

\section{Numerical results}
\label{sec:numerical}

We are now ready to describe the numerical evaluation  of 
the contributions  of the Set I(j) to the electron $g\!-\!2$.
The largest contribution,
which is mass independent,  
comes from the case where both fermion loops consist of electrons. 
We first made a preliminary evaluation of the coefficient of $(\alpha/\pi)^5$
by VEGAS \cite{Lepage:1977sw} using relatively small sampling points.
The results may be summarized as follows:
\begin{align} 
A_1^{(10)} (\text{Set I(j): Method~B}) 
           &= -0.072~8843~8~(57)~+~0.0732~732~(118) \nonumber \\
                                    &= ~0.000~388~9~(131),  \\
A_1^{(10)} (\text{Set I(j): Method~C1}~~g^{\mu \nu}) 
           &= -0.059~537~(39) ~+~ 0.059~917~ (30)    \nonumber \\ 
                                    &=~0.000~380~(50), \\
A_1^{(10)} (\text{Set I(j): Method~C1}~~q^\mu q^\nu)
           &= -0.0161~450~(58) ~+~ 0.016~432~ (161)    \nonumber \\ 
                                    &=~0.000~287~(171), \\
A_1^{(10)} (\text{Set I(j): Method~C2}) 
           &= -0.047~895~3~(60)~+~0.048~290~3~ (62) \nonumber \\     
                                    &=~0.000~395~0~(87) ,
\end{align}
where the first and second terms on the first line of each case
are from \LLp\ and \LLc\, respectively.
Their vacuum-polarization functions are obtained from the terms
proportional to $g^{\mu \nu}$ in Method~B, Method~C2, and the first Method~C1,
and the terms proportional to $q^\mu q^\nu$ in the second Method~C1.
Method~B and Method~C2 were evaluated on {\sl hp}'s Alpha station.
For Method~B,  we used  $10^8$ points per iteration  and
250 iterations, followed by $10^9$ points per iteration and 160 iterations
for VEGAS integration.
For  Method~C2, we used $10^8$ points per iteration and 100 iterations.
Both of  Method~C1 were carried out  with  
$10^6$ points per iteration and 100 iteration on a PC with Intel's Core 2 processor.
The Method~B requires more sampling statistics for VEGAS than the Method C
in order to reduce the uncertainty to the level of the latter,
primarily because the renormalization is carried out
by point-by-point cancellation between divergent pieces 
of the integrand.
The individual terms of Method~B, Method~C1, and Method C2  
are different from each other
because of different treatments of renormalization and routing selection.
Note also that the $g^{\mu \nu}$ term and $q^\mu q^\nu$ term in Method~C1 
give different integrands.
The good agreement of these four cases within the numerical uncertainties 
provides a strong assurance of correctness of all  calculations.

The most prominent feature of these calculations is that
the \LLp\ and \LLc\ parts are nearly equal in magnitude
and almost cancel each other.
In view of similar analytic structures of these integrals, 
this suggests the possibility that cancellation takes place
not only between the integrals as a whole but also between
the integrands at many points in the domain of integration.
If this is the case, one should be able to reduce
the calculated uncertainty significantly by performing
integration of the combination $2 \times \LLp + 4 \times \LLc$.
In order to verify this conjecture, we have carried out an extensive 
computation of the combination in both Method B and Method C1.
  
The production job for evaluation of combined
\LLp\ and \LLc\ were carried out 
on RIKEN's Super Combined Cluster System (RSCC) with 128 or 256 processors.  
It turns out that the vacuum-polarization functions $\Pi(q^2)$ obtained from
$g_{\mu \nu}$ term  and $ q^\mu q^\nu$ term have  analytically different
structures in Method C1 and C2 even after  combining \LLp\ and \LLc.
This provides us with more opportunity to
check our  calculation. The numerical results obtained 
by various methods are summarized in Table~\ref{tab:e-results}.

\begin{table}
\begin{ruledtabular}
\begin{tabular}{ccdcc} 
 method   & $g^{\mu \nu}$  or $q^\mu q^\nu$   & \multicolumn{1}{r}{contribution~~~} &  sampling  points & iteration \\
          &                &                      &  per iteration    &          \\ 
\hline 
B  &  $g^{\mu \nu}$  &   0.000~396~4~(59) &   $10^8$, $10^{10}$  &  50, 100 \\
C1 &  $g^{\mu \nu}$  &   0.000~398~2~(31) &   $10^9$             &  101       \\
C1 &  $q^\mu  q^\nu$ &   0.000~393~8~(88) &   $10^9$             &  50  \\
C2 &  $g^{\mu \nu}$  &   0.000~397~6~(25) &   $10^9$             &  80  \\

\end{tabular} 
\caption{ 
A contribution to the mass-independent term of electron $g\!-\!2$, $A_1^{(10)}$, 
from the diagrams of Set I(j) calculated in various methods. All methods 
are analytically independent.
The overall factor $\left(\frac{\alpha}{\pi}\right)^5$ is omitted for simplicity.
The second column shows from which term, $g^{\mu \nu}$ or $q^\mu q^\nu$ term,  the vacuum-polarization function $\Pi(q^2)$ is obtained. 
The numeral in the parenthesis stands for the uncertainty 
in the last two digits.  All calculations were carried out on RSCC. 
} 
\label{tab:e-results}
\end{ruledtabular}
\end{table}

Four values listed in Table~\ref{tab:e-results} are independent of each other. 
Thus, combining these results statistically, we obtain 
\begin{equation} 
A_1^{(10)} ( \text{Set I(j): combined} ) = 0.000~397~5~(18)
\label{eq:mass-indep-result}
\end{equation}
as the best estimate of the term $A_1^{(10)} ( \text{Set I(j)})$.

The mass-dependent
contributions to the electron $g\!-\!2$ involving muons and/or tau leptons are
also calculated using the combined programs of Method~C1 with 
$g^{\mu\nu}$ term and with $q^\mu q^\nu$ term.
Singular behavior of the integral caused by heavier leptons makes 
convergence of the integrand rather difficult.
But, the contributions  themselves are very small and currently of no interest 
compared with the experimental uncertainty. Therefore we do not need the 
precise values of the mass-dependent contribution.
They are summarized in Table~\ref{tab:e-mass-dep-results} \cite{Watanabe:2008}.

Summing up all mass-dependent terms and the mass-independent
contribution Eq.~(\ref{eq:mass-indep-result}), we find
the total contribution to the electron $g\!-\!2$ from Set I(j) given in
Eq.~(\ref{eq:electron-final-result}).

\begin{table}[t]
\begin{ruledtabular}
\begin{tabular}{ccdc} 
 loop fermions   & $g^{\mu\nu}$ or $q^\mu q^\nu$  & \multicolumn{1}{r}{contribution~~~~~~~~~~}  &  iteration \\ 
 \hline 
 $(e,\,\mu)$     & $g^{\mu\nu}$ & 2.281~(60) \times 10^{-6}  & $51$ \\  
 $(e,\,\mu )$    & $q^\mu q^\nu$& 2.290~(115)\times 10^{-6}  & $60$ \\  
 $(\mu,\,\mu)$   & $g^{\mu\nu}$ & 1.185~(17) \times 10^{-7}  & $145$ \\   
 $(\mu,\,\mu)$   & $q^\mu q^\nu$& 1.284~(4)  \times 10^{-7}  & $55$ \\   
 $(e,\,\tau)$    & $g^{\mu\nu}$ & 1.332~(95) \times 10^{-8}  & $100$ \\  
 $(e,\,\tau)$    & $q^\mu q^\nu$& 1.62~~(51) \times 10^{-8}  & $50$ \\  
 $(\mu,\,\tau)$  & $g^{\mu\nu}$ & 4.988~(27) \times 10^{-9}  & $50$ \\   
$(\mu,\,\tau)$   & $q^\mu q^\nu$& 5.007~(55) \times 10^{-9}  & $50$ \\
$(\tau,\,\tau)$  & $g^{\mu\nu}$ & 4.008~(99) \times 10^{-10} & $90$ \\   
$(\tau,\,\tau)$  & $q^\mu q^\nu$& 4.541~(13) \times 10^{-10} & $50$ \\     
\end{tabular} 
\caption{ 
Mass-dependent contributions to the electron $g\!-\!2$ from 
the diagrams of Set I(j) with $e$, $\mu$ and/or $\tau$ lepton loops.
All integrands are constructed in Method C1.  
The overall factor $\left(\frac{\alpha}{\pi}\right)^5$ 
is omitted for simplicity. The number of sampling points 
per iteration for VEGAS integration is $10^9$ for all calculations.
The numeral in the parenthesis stands for the uncertainty 
in the last two digits.  All calculations were conducted on RSCC.
} 
\label{tab:e-mass-dep-results}
\end{ruledtabular}
\end{table}

We also present the contributions to the muon $g\!-\!2$. They 
were calculated by replacing the
external electron by a muon in the combined program of Method~C1 and/or C2.
They are listed in Table~\ref{tab:muon-results}.
The dominant contribution arises when both of light-by-light scattering loops 
consist of electrons.  Statistically combining three results listed in 
Table~\ref{tab:muon-results} of this contribution,
we found
\begin{equation}
A_2^{(10)}(m_\mu/m_e)( \text{Set I(j)}_{(e,e)}: \text{combined} ) = -1.247~26~(12)~,
\label{eq:eee-result}
\end{equation}
where the subscript $(e,e)$ implies that both fermion loops consist of electrons.
Including all other contributions, we found the mass-dependent contribution to
the muon $g\!-\!2$ given in Eq.~(\ref{eq:muon-final-result}).
By using the asymptotic expansion of the vacuum-polarization function with
respect to the transfer momentum, 
Kataev obtained the leading terms of this contribution: 
\cite{Kataev:1991az,Kataev:1991cp} 
\begin{align}
&A_2^{(10)}(m_\mu/m_e)( \text{Set I(j)}_{(e,e)}: \text{asympt.} ) 
\nonumber \\
&~~~~~~=-\frac{ 1}{2 } a_4^{[2,l-l]} + 
   \left [ \ln \left( \frac{ m_\mu}{m_e} \right ) -\frac{5}{4} \right ]
         \left[ \frac{11}{36} - \frac{2}{3} \zeta(3) \right ] 
    + \mathcal{O} \left( \frac{m_e}{m_\mu} \right )
\nonumber \\
&~~~~~~=-\frac{1}{2} a_4^{[2,l-l]} - 2.0237 + \mathcal{O} \left( \frac{m_e}{m_\mu} \right ),
\label{eq:asympt-result}
\end{align}
where $a_4^{[2,l-l]}$ is the unknown constant term of 
the asymptotic expansion of the eighth-order
vacuum-polarization function formed by two light-by-light subdiagrams.
Comparing our result (\ref{eq:eee-result}) and the formula (\ref{eq:asympt-result}), 
we obtain
\begin{equation}
a_4^{[2,l-l]} = -1.5529 + \mathcal{O} \left( \frac{m_e}{m_\mu} \right )~.
\end{equation}

\begin{table}[t]
\begin{ruledtabular}
\begin{tabular}{cccdc} 
 loop fermions  &  method &   $g^{\mu\nu}$ or $q^\mu q^\nu$ &\multicolumn{1}{r}{contribution~~~~} &  iteration \\ 
 \hline 
 $(e,\,e)$         & C1 & $q^\mu q^\nu$ &  -1.247~28~(25)     & 50 \\ 
 $(e,\,e)$         & C1 & $g^{\mu \nu}$ &  -1.247~30~(15)     & 71 \\ 
 $(e,\,e)$         & C2 & $g^{\mu \nu}$ &  -1.247~08~(31)     & 25 \\      
 $(e,\,\mu)$       & C2 & $g^{\mu \nu}$ &  -0.016~455~(71)      & 20  \\
 $(e,\,\tau)$      & C2 & $g^{\mu \nu}$ & 0.109~9~(51)\times 10^{-3} & 20  \\
 $(\mu,\,\tau)$    & C2 & $g^{\mu \nu}$ & 0.149~0~(13)\times 10^{-3} & 20  \\
 $(\tau,\,\tau)$   & C2 & $g^{\mu \nu}$ & 0.189~3~(5) \times 10^{-4}  & 20 
\end{tabular} 
\caption{ 
Contributions to the mass-dependent term of  muon $g\!-\!2$ from 
the diagrams of Set I(j) with $e$, $\mu$ and/or $\tau$ lepton loops.  
The overall factor $\left(\frac{\alpha}{\pi}\right)^5$ 
is omitted for simplicity. The number of sampling points 
per iteration for VEGAS integration is $10^9$ for all calculations.
The numeral in the parenthesis stands for the uncertainty 
in the last two digits.  
All calculations were carried out on RSCC.
} 
\label{tab:muon-results}
\end{ruledtabular}
\end{table}

\section{discussion}
\label{sec:conclusion}

In this paper, we report the evaluation of the contribution 
to the electron $g\!-\!2$ and muon $g\!-\!2$ from Set I(j)   
which consists of six vacuum-polarization diagrams  
formed by two light-by-light scattering subdiagrams. 
The contribution to the electron $g\!-\!2$ given in 
Eq.~(\ref{eq:electron-final-result}) amounts $ 0.000~027\times 10^{-12}$, 
which is far smaller than the current experimental uncertainty $ 0.28 \times 10^{-12}$
in Eq.~(\ref{eq:aHV08}). 
Thus far we have
 no clear explanation of why the contribution from Set I(j) is so small
compared with other diagrams of the tenth order.

We have also demonstrated 
the utility of the Ward-Takahashi identity 
to deal with the
light-by-light scattering subdiagram. 
Although it may not always help us to streamline 
the work-flow for writing  numerical programs, 
we continue to examine its application to 
the computation of the tenth-order diagrams containing 
a light-by-light scattering subdiagram 
that have not been evaluated yet.

\begin{acknowledgments} 
This work is supported in part by JSPS Grant-in-Aid 
for Scientific Research (C) 19540322.
M.~H. is also supported in part by JSPS Grant-in-Aid 
for Scientific Research (C) 20540261. 
T.~K.'s work is supported by the U.~S.~National Science 
Foundation under Grant PHY-0355005. 
T.~K. also thanks 
JSPS Invitation Program for Research in Japan S-07165, 2007.
We thank  Dr.~A.~L.~Kataev for reminding us of his analytic work.
Numerical computations were mostly conducted on 
the RIKEN Super Combined Cluster System (RSCC). 
A part of preliminary computations was also conducted  
on the computers of the theoretical particle-physics group (E-ken), Nagoya University. 
\end{acknowledgments}

\appendix

\section{Vacuum-polarization insertion into the anomaly integral}
\label{AppendixA}

Analyticity of a  vacuum-polarization function $\Pi(q^2)$ ensures that
once-subtracted dispersion relation between its real part and imaginary part is given by
\begin{equation}
\frac{ \rm{Re} \Pi(q^2)}{q^2} = \frac{1}{\pi} \int^\infty_{0}    dk^2 
\frac{ \rm{Im} \Pi(k^2)}{ k^2 (k^2 - q^2) }~.
\label{DR}
\end{equation}
In the cases of vacuum-polarization
of second, fourth, and sixth orders, the cut starts at $q^2=4m^2$, where $m$ is the
electron mass,  and the
imaginary part of $\Pi(q^2)$ is nonvanishing for $q^2 > 4m^2$ only. 
For $\Pi (q^2)$ of Set I(j), however, the cut starts
at $q^2=0$ because of three photon intermediate states,
and is included in Eq.~(\ref{DR}). 
This is a novel feature encountered for the first time in the eighth-order
vacuum polarization.
Eq.~(\ref{DR}) also assumes that $\Pi (q^2)$ has no pole at $q^2 =0$.
This may be justified as follows:
At the threshold $q^2=0$ the absorptive part of $\Pi(q^2)$
is proportional to the square of light-by-light-scattering amplitude,
which is proportional to $q^8$ because the light-by-light amplitude
is known to be proportional to $q^4$ \cite{Euler:1936}.
Meanwhile, three photon propagators could produce $1/q^6$ at most
so that $ \rm{Im} \Pi (q^2)$ behaves as $q^2$ or 
even higher positive  power of $q^2$ as $q^2 \rightarrow 0$.
Thus the singularity at $q^2=0$ cannot be a pole.

Eq.~(\ref{DR}) guarantees that $\Pi(q^2)$
can be expressed by a spectral representation 
\begin{equation}
\frac{\Pi(q^2)}{q^2} = \int_0^\infty dk^2 \frac{ \rho(k^2)}{ -q^2 + k^2},
\label{eq:spectral-rep} 
\end{equation}
where
\begin{equation}
  \rho (k^2) = \frac{1}{\pi}\frac{ \rm{Im} \Pi (k^2)  }{k^2 }~.
\end{equation} 
The effect of inserting a vacuum-polarization diagram into a photon line with momentum $q$ is
thus obtained by replacing the photon propagator by a 
sum of massive vector propagators
whose mass squared is $k^2$:
\begin{equation}
\frac{1}{q^2} \longrightarrow \frac{-\Pi(q^2) }{q^2} = \int_0^{\infty} dk^2 \frac{\rho(k^2)}{q^2 - k^2}~.
\end{equation}
This is easily translated into a Feynman-parametric integral formula.
If a vacuum-polarization is inserted into a photon line $z_a$, we need
to replace a photon mass $\lambda^2$ by $k^2$,
multiply the spectral function $\rho(k^2)$ to the whole integrand,
and integrate over a ``photon mass''  $k^2$. This is the most efficient
way to describe an effect of the vacuum-polarization insertion in the
anomaly calculation.
 
The spectral function, or the imaginary part of $\Pi(q^2)$, however,
is not always available, especially in higher-order cases \cite{Kinoshita:1979dy}.
On the other hand, we can directly construct $\Pi(q^2)$
itself, or its real part, in the Feynman-parameter
space using the Feynman-Dyson rules.
 Thus our problem becomes how to express
the effect of vacuum-polarization insertion by using the real
part of $\Pi(q^2)$.

Let us specifically consider the anomaly contribution from a diagram 
in which a vacuum-polarization diagram is inserted into
the second-order vertex diagram.  We will omit the overall factor $\alpha/\pi$ 
for simplicity and  set the electron mass $m$ to unity.
It is given as the integral over the Feynman parameters 
\cite{Kinoshita:1990}
\begin{equation}
M_{2,P} = \int_0^\infty dk^2 \rho(k^2) 
\int \frac{(dz)}{U^2} \frac{F_0}{V+  z_a k^2},
\label{eq:M2PwRho}
\end{equation} 
where a Feynman parameter assigned to the photon line is $z_a$ and
that to the fermion line is $z_1$. 
The explicit form of $F_0$, $V$,$\cdots$, etc. are \cite{Kinoshita:1990}  
\begin{align}
&(dz) = dz_1 dz_a \delta(1-z_1-z_a), \quad  U=1,  \quad A_1=1-\frac{z_1}{U} \nonumber \\
&F_0 = z_1 A_1(1-A_1), \quad V = z_1 - z_1 A_1 + z_a\lambda^2 ~.
\label{M2explicit}
\end{align}
Comparing Eq.~(\ref{eq:M2PwRho}) to Eq.~(\ref{eq:spectral-rep}), we find 
\begin{equation}
M_{2,P}= \int \frac{(dz)}{U^2} \frac{F_0 }{V} ( - \Pi(q^2))
\label{eq:M2PwPi}
\end{equation}
with 
\begin{equation}
q^2 = - \frac{V}{z_a}~.
\end{equation}    
Substituting the explicit forms in Eq.~(\ref{M2explicit}) and identifying
$z_1=y$ and $z_a=1-y$, one find that Eq.~(\ref{eq:M2PwPi}) becomes
Eq.~(\ref{eq:fromBroadhurst}).

Higher-order diagrams contributing to the magnetic moment
  may have $V^2$ or higher powers of $V$ in the denominators.  
We can easily extend the above method to such cases.
Namely, the effect of vacuum-polarization insertion into 
a photon line $z_a$ is expressed by replacing the denominator $1/V^n$
for $n \ge 1$ according to a following rule:
\begin{equation}
\frac{1}{V^{n}} \longrightarrow (-1)^{n}\frac{1}{(n-1)!}
                                \frac{\partial^{n-1}}{\partial V^{n-1}} 
                                      \left ( \frac{\Pi(-V/z_a)}{V} \right )~.
\end{equation}
This rule is used in our forthcoming papers dealing with the insertion
of vacuum-polarization loops in the magnetic moments of
fourth, sixth, and eighth orders.

\section{Structure of the integrand of Method C}
\label{AppendixB}

The integral Eq.~(\ref{eq:Pi_parametricIntegral}) 
generated by Method C is very lengthy containing more than
30,000 terms.  The integrand, however, can be shortened by
observing its structure carefully.
When the term proportional to $g^{\mu \nu}$ is projected
out from Eq.~(\ref{integformulaC}),    
$D_i^\mu$ and $D_j^\nu$ in $\mathbb{D}^{\mu \nu}$ 
must be ``contracted'' with other $D_k$\cite{Kinoshita:1990}.  
Knowing it, we can organize  $H_{(r),\,h}$ in the form 
\begin{align} 
 H_{(r),\,h}
 &= 
 \sum_{1 \le k < l \le 8}   
  T_{(r),\,h}^{kl} 
 \left\{  
  \left( 
   \sum_{i=1}^4 d_i z_i B^{\prime}_{i k} 
  \right) 
  \left( 
   \sum_{j=5}^8 d_j z_j B^{\prime}_{j l} 
  \right) 
  + (k \leftrightarrow l) 
 \right\} \nonumber \\ 
 & 
 + 
 F_{(r),\,h}  
 \sum_{i=1}^4 d_i z_i  
 \sum_{j=5}^8 d_j z_j \,B_{ij}  
 \, ,  \label{eq:artifice}
\end{align}
where $T_{(r), \, h}$ and $F_{(r), \, h}$
are expressed in terms of 
``building blocks'' $B_{ij}$, $z_i$, and $A_i$ \cite{Aoyama:2005kf}.
Then, 
the number of arithmetic operations is dramatically  reduced
and the computational time becomes less than one tenth
of the program without the above artifice.

\bibliographystyle{apsrev}
\bibliography{b}

\end{document}